\documentclass[12pt]{article}
\usepackage{multirow}
\usepackage{amsthm}
\usepackage{amssymb}
\usepackage{natbib}
\usepackage{amsmath,color}
\usepackage{mathrsfs}  
\usepackage[ruled,vlined]{algorithm2e}
\usepackage{bbold}
\usepackage{bbm}
\usepackage{multirow}
\usepackage[table,xcdraw]{xcolor}
\usepackage{subcaption}
\usepackage[utf8]{inputenc}
\usepackage[export]{adjustbox}
\usepackage{wrapfig}
\usepackage[normalem]{ulem}
\usepackage{etoolbox}
\usepackage{enumerate}
\usepackage{diagbox}
\usepackage{booktabs}
\usepackage[toc,page]{appendix}
\usepackage{comment}

\def\RR{\mathbbm{R}}
\def \EE {\mathbbm E}

\def \diff {\mathrm{diff}}

\def \Var {\mathrm{V a r}}

\def \Cov {\mathrm{ C o v}}

\def \X {\mathbf{X}}
\def \Y {\mathbf{Y}}
\def \Z {\mathbf{Z}}

\def \W {\mathbf{W}}

\def \G {\mathbf{G}}
\def \D {\mathbf{D}}

\def \bSigma {\mathbf{\Sigma}}

\def \bu {\mathbf{u}}
\def \bv {\mathbf{v}}

\def \df{{\rm df}}
\def\trans{^{\scriptscriptstyle \sf T}}

\def \tW{\widetilde{W}}

\def \bSigma{\boldsymbol{\Sigma}}

\newcommand{\simiid}{\overset{\rm i.i.d}{\sim}}

\def\titlefull{{MATES: Multi-view Aggregated Two-Sample Test}}
\RequirePackage[colorlinks,
            linkcolor=blue,
            anchorcolor=blue,
            citecolor=blue
            ]{hyperref}
\usepackage{cleveref}
\newtheorem{theorem}{Theorem}
\newtheorem{lemma}{Lemma}

\newtheorem{remark}{Remark}

\crefname{enumi}{}{}
\Crefname{enumi}{}{}
\crefname{equation}{}{}
\Crefname{equation}{}{}
\crefname{lemma}{}{}
\Crefname{lemma}{Lemma}{Lemmas}
\crefname{theorem}{}{}
\Crefname{theorem}{Theorem}{Theorems}
\Crefformat{appendix}{Supplementary~#2#1#3}
\usepackage{enumitem}

\usepackage{authblk}

\begin{document}
\def\spacingset#1{\renewcommand{\baselinestretch}%
{#1}\small\normalsize} \spacingset{1}

\def\titlefull{MATES: Multi-view Aggregated Two-Sample Test}
\title{\Large\bf \titlefull \vspace{0.5em}}
\author[1,*]{Zexi Cai}
\author[1,*]{Wenbo Fei}
\author[2,*,**]{Doudou Zhou}
\affil[1]{Department of Biostatistics, Columbia University, New York, USA}
\affil[2]{Department of Statistics and Data Science, National University of Singapore, Singapore}
\affil[*]{Alphabetical order}
\affil[**]{Corresponding author: ddzhou@nus.edu.sg}
\date{}
\maketitle

\begin{abstract}
The two-sample test is a fundamental problem in statistics with a wide range of applications. In the realm of high-dimensional data, nonparametric methods have gained prominence due to their flexibility and minimal distributional assumptions. However, many existing methods tend to be more effective when the two distributions differ primarily in their first and/or second moments. In many real-world scenarios, distributional differences may arise in higher-order moments, rendering traditional methods less powerful. To address this limitation, we propose a novel framework to aggregate information from multiple moments to build a test statistic. Each moment is regarded as one view of the data and contributes to the detection of some specific type of discrepancy, thus allowing the test statistic to capture more complex distributional differences. The novel \textbf{m}ulti-view \textbf{a}ggregated two-sample \textbf{tes}t (MATES) leverages a graph-based approach, where the test statistic is constructed from the weighted similarity graphs of the pooled sample. 
Under mild conditions on the multi-view weighted similarity graphs, we establish theoretical properties of MATES, including a distribution-free limiting distribution under the null hypothesis, which enables straightforward type-I error control. Extensive simulation studies demonstrate that MATES effectively distinguishes subtle differences between distributions. We further validate the method on the S\&P100 data, showcasing its power in detecting complex distributional variations.
\end{abstract}

\noindent%
\vfill

\newpage
\spacingset{1.65} %

\allowdisplaybreaks

\section{Introduction}\label{sec:intro}

The two-sample test is a classical statistical problem with broad applicability across many fields, including genomics, finance, and biomedical research. Its objective is to determine whether two samples originate from the same underlying distribution. Formally speaking, for two independent samples $\X_1, \ldots, \X_m \simiid F_{\X}$ and $\Y_1, \ldots, \Y_n \simiid F_{\Y}$, one would like to test $H_0: F_{\X} = F_{\Y}$ against $H_1: F_{\X} \neq F_{\Y}$.

For univariate data, a variety of classical parametric tests are available to assess differences between distributions in specific aspects. For example, Hotelling $T^2$ test \citep{hotelling1931generalization} and Welch's $t$-test \citep{welch1947generalization} are commonly used to detect mean differences, while Bartlett's test \citep{bartlett1937properties} and Levene's test \citep{levene1960robust}  focus on  testing variance differences. %
Nonparametric methods, which avoid strict parametric assumptions, are particularly useful for testing general distributional differences. Examples include the Kolmogorov-Smirnov test \citep{kolmogorov1933sulla}, the Cram\'er-von Mises test \citep{cramer1928composition}, and the Anderson-Darling test \citep{anderson1952asymptotic}, which leverage empirical distribution functions, and rank-based tests like the Mann-Whitney U test \citep{mann1947test} and the Brunner-Munzel test \citep{brunner2000nonparametric}.

In high-dimensional data analysis, parametric methods often face limitations due to the curse of dimensionality. As a result, many nonparametric tests are developed to address this issue. For instance, tests for detecting differences in high-dimensional means include \cite{chen2010two, tony2014two, xu2016adaptive, zhang2020simple, zhang2021more, jiang2024nonparametric}. 
Similarly, \cite{chen2010tests, li2012two, cai2013two, li2014hypothesis, srivastava2014tests, chang2017comparing} developed methods for assessing differences in high-dimensional covariances.

Beyond mean and covariance comparisons, recent decades have seen a wide range of methods aiming at detecting broader distributional differences in high-dimensional settings.  These include classification-based tests \citep{lopez2016revisiting,kirchler2020two,kim2021classification,hediger2022use}, distance-based tests \citep{szekely2013energy,biswas2014nonparametric,li2018asymptotic}, kernel-based tests \citep{gretton2008kernel,gretton2009fast,gretton2012kernel,song2023generalized}, graph-based tests \citep{friedman1979multivariate,schilling1986multivariate,henze1988multivariate,rosenbaum2005exact,chen2013graph,chen2017new,bai2023robust}, and rank-based tests \citep{baumgartner1998nonparametric,hettmansperger1998affine,pan2018ball,deb2023multivariate,zhou2023new}. More recently, several efforts were made to improve distance and kernel-based methods, including \cite{shekhar2022permutation,schrab2023mmd,makigusa2024two} that generalized the maximum mean discrepancy (MMD)-based test, and \cite{zhu2021interpoint,chakraborty2021new,yan2023kernel,chatterjee2024boosting} that extended and further unified the MMD and energy distance-based tests, and \cite{liu2020learning} that utilized deep learning techniques to learn the optimal kernel for the kernel-based two-sample tests.

While these methods have demonstrated strong performance in many applications, they are generally more effective for detecting mean and covariance differences between two samples in the high-dimensional settings, and less sensitive to variations in the higher-order moments. %
To illustrate this, consider two samples $\X_i = (X_{i1}, \ldots, X_{id})$ and $\Y_i = (Y_{i1}, \ldots, Y_{id})$, where $X_{ij} \simiid t_{15}$ and $Y_{ij} \simiid N(0,15/13),j=1,\ldots,d$, with $m=n=50$ and $d=200$. The two distributions have the same mean and covariance, but the first distribution has heavier tails than the second. To assess the performance of existing methods, we evaluate $12$ commonly used or recently proposed two-sample tests:
Rosenbaum's cross-matching test (CM, \cite{rosenbaum2005exact}), 
the generalized edge count test (GET, \cite{chen2017new}), 
the ball divergence-based test (BD, \cite{pan2018ball}), 
the kernel-based test with generalized energy distance (GED, \cite{chakraborty2021new}), 
the random forest-based test (RF, \cite{hediger2022use}), 
the multivariate rank-based test using measure transportation (MT, \cite{deb2023multivariate}), 
the generalized kernel tests (GPK, \cite{song2023generalized}), 
the rank in similarity graph edge-count test (RISE, \cite{zhou2023new}), 
and the MMD test and its variants, including 
the kernel-based MMD test (MMD, \cite{gretton2012kernel}), 
the permutation-free MMD test (xMMD, \cite{shekhar2022permutation}), 
the aggregated MMD test (aMMD, \cite{schrab2023mmd}),
and 
the Mahalanobis aggregated MMD test with multiple kernels (mMMD, \cite{chatterjee2024boosting}). 
Using a significance level of $\alpha = 0.05$ and $1000$ repetitions, \Cref{tab:example} shows the empirical power of these methods. Most tests exhibit power close to the significance level, highlighting their limited ability to detect differences when the two distributions %
{differ primarily} in higher-order moments. %
{Extensive numerical experiments in \Cref{sec:simulation} further confirm the issue under various scenarios.}

\begin{table}[!htbp]
    \centering\spacingset{1}
    \caption{Empirical power of $12$ two-sample tests on the example data with $\alpha = 0.05$.}
    \label{tab:example}
    \resizebox{\textwidth}{!}{%
    \begin{tabular}{|c|c c c c c c c c c c c c|}
    \hline 
       Method & CM & GET & BD & GED & RF & MT & GPK & RISE & MMD & xMMD & aMMD & mMMD \\\hline 
       Power & 0.05 & 0.06 & 0.05 & 0.09 & 0.07 & 0.06 & 0.04 & 0.06 & 0.05 & 0.06 & 0.06 & 0.01
        \\\hline 
    \end{tabular}
    }
\end{table}

This phenomenon may stem from the concentration of Euclidean distance in high-dimensional settings \citep{hall2005geometric}. Specifically, for high-dimensional points $\X, \Y \in \RR^d$, suppose that $\lim_{d \rightarrow \infty}E\|\X-E \X\|_2^2/d  = \sigma_1^2$, $\lim_{d \rightarrow \infty}E\|\Y-E \Y\|_2^2/d  = \sigma_2^2$, and $\lim_{d \rightarrow \infty}\|E \X- E \Y\|_2^2/d  = \upsilon^2$. Then, as $d \rightarrow \infty$, the pairwise distances within and between samples, scaled by $1 / \sqrt{d}$, converge to $\sqrt{2} \sigma_1$, $\sqrt{2} \sigma_2$, and $\sqrt{\sigma_1^2 + \sigma_2^2 + \upsilon^2}$, respectively. When $\sigma_1 = \sigma_2$ and $\upsilon = 0$,
the within- and between-sample distances become nearly indistinguishable, causing many tests utilizing Euclidean distance to struggle in differentiating the samples. We provide more insights into this phenomenon in the discussion of \Cref{thm:decomposition} later.

Higher-order moments, inherently, carry crucial information in many real-world scenarios, especially when the first two moments of the distributions are similar. For instance, in finance, stock returns rarely follow a normal distribution, and investors often favor positive skewness and avoid high kurtosis \citep{scott1980direction, de2022investigation}. Extensions to the traditional Markowitz portfolio theory, which focuses on mean and variance only \citep{markowitz1952portfolio}, have thus been developed to incorporate higher moments \citep{kraus1976skewness, lassance2021portfolio, khashanah2022we}. Despite the importance of higher-order moments, most existing two-sample tests do not fully leverage them. This underscores the need for tests that utilize information from multiple perspectives and are capable of detecting differences in higher moments.

To this end, \cite{chakraborty2021new} introduced the generalized energy distance test (GED), which achieves the best performance among the methods in Table \ref{tab:example}. Their study showed that in high-dimensional settings, tests relying on the standard Euclidean energy distance are limited to detecting differences in mean and covariance trace. Therefore, they proposed a modified energy distance metric to capture higher-order differences that standard metrics miss. However, as illustrated in our example, even GED shows limited power of  $0.09$. This result indicates that simply replacing the Euclidean distance by other metrics may not be sufficient and efficient to detect higher-moment discrepancies, thus emphasizing the necessity for further advancements in two-sample testing methods to more effectively identify complex distributional variations.

As a result, we develop a more flexible and comprehensive framework for testing two-sample differences that can incorporate multiple moment information as a special case. In fact, each moment can be regarded as some specific characteristic of the data, which is termed a \textit{view}. Considering a single view of the data may miss some subtle yet important information from the two samples, such as higher-order moments. In this paper, we propose a new graph-based \textbf{m}ulti-view \textbf{a}ggregated two-sample \textbf{tes}t (MATES) for modern data. Here, the multiple views in MATES refer to various aspects of the differences between the two samples, which are summarized by varied choices of distance measures, similarity graphs, and edge weights. We construct the test statistic by aggregating information from these views, enabling MATES to capture complex distributional differences that traditional methods may overlook. In our example in \Cref{tab:example}, MATES achieves the power of $\mathbf{0.91}$, demonstrating its effectiveness. As will be shown in \Cref{sec:simulation}, under various dimension and distribution settings, MATES is also robust and outperforms other competing methods. Under mild conditions on the multi-view weighted similarity graphs, we show that MATES possesses a distribution-free limiting distribution under the null hypothesis, enabling straightforward control of type-I error. 

The MATES method also performs well in real data applications, where the difference between two samples may be more obvious in their higher-order moments instead of only in mean or variance. In the S\&P100 dataset analyzed in \Cref{sec:realdata}, we observe that the release of ChatGPT had an impact on the stock returns of large-cap companies, a change reflected in the higher-order moments that signal potential rare events. This effect aligned with the notable rise in stock prices for Nvidia and other ``Magnificent 7'' companies starting in January 2023. %

The rest of the paper is organized as follows. \Cref{sec:method} presents the details of the new test, and \Cref{sec:simulation} demonstrates the superior performance of the proposed test in distinguishing subtle distributional differences compared to existing alternatives. \Cref{sec:theory} provides the theoretical properties of the test. We apply the proposed method to the S\&P100 data in \Cref{sec:realdata} and conclude the paper with a discussion in \Cref{sec:discussion}. The proofs of the theorems and other supporting information are deferred to the Supplement Material.

\section{Method}\label{sec:method}
We now introduce the general formulation of our proposed test statistic, which utilizes similarity graphs constructed from the pooled sample. Let $\Z = (\Z_1, \ldots, \Z_N)$ represent the pooled sample, where $N = m + n$, with $\Z_i = \X_i$ for $i = 1, \ldots, m$ and $\Z_{m+j} = \Y_j$ for $j = 1, \ldots, n$. Our method captures multiple views of the data, organized across three levels: similarity measures, similarity graphs, and edge weights. These views provide complementary perspectives on the data, enabling the detection of complex distributional differences that traditional methods may miss.

\textbf{Similarity Measures.} Selecting an appropriate similarity or dissimilarity measure is essential as it enables each view to capture distinct aspects of the sample distributions, thereby enhancing the test's ability to detect subtle differences. Let $\D^{(s)} = [D_{ij}^{(s)}]_{i,j=1}^N \in \RR^{N\times N}$ represent the dissimilarity matrix for the $s$th view of the pooled sample. Different similarity or distance measures can be used to capture various distributional characteristics. For example, in the $s$th view for Euclidean data, we may use the Manhattan distance based on the $s$th moment for multivariate data, $
    D_{ij}^{(s)} = \sum_{r=1}^{d} |Z_{ir}^{s} - Z_{jr}^{s}|,
$ or the $\ell_s$-distance, 
$
    D_{ij}^{(s)} = (\sum_{r=1}^{d} |Z_{ir} - Z_{jr}|^{s})^{1/s}
$.
With this choice, each view is thus tailored to capture specific moment-based differences between samples. We also discuss the construction of dissimilarity matrices for non-Euclidean data in \Cref{rem1}.

\textbf{Similarity Graphs.} While distance-based methods provide an intuitive approach to capturing differences between samples, they are often sensitive to outliers and may require certain moment conditions for theoretical validity. In contrast, graph-based methods offer robustness to outliers and heavy-tailed distributions \citep{friedman1979multivariate,schilling1986multivariate,rosenbaum2005exact,chen2013graph,chen2017new}. To leverage this robustness, we propose constructing similarity graphs, denoted as $\G^{(s)} = [G_{ij}^{(s)}]_{i,j=1}^N$, based on the dissimilarity matrix $\D^{(s)}$. 

Common similarity graphs include the $k$-nearest neighbor graph ($k$-NNG), the $k$-minimum spanning tree ($k$-MST, \citet{friedman1979multivariate}), the $k$-minimum distance non-bipartite pairing ($k$-MDP, \citet{rosenbaum2005exact}), and the recently proposed robust $k$-nearest neighbor graph (r$k$-NNG, \citet{zhu2023robust}). For example, in the $k$-NNG, $G_{ij}^{(s)} = 1$ if observation $j$ is among the $k$ nearest neighbors of observation $i$ based on the dissimilarities in $\D^{(s)}$, and $G_{ij}^{(s)} = 0$ otherwise. 

\textbf{Edge Weights.} 
To capture more detailed relationships between observations, the similarity graph $\G^{(s)}$ can use various edge weights instead of binary connection status only (i.e., $0$ for unconnected and $1$ for connected). Hence, we propose to use the weighted graph denoted by $\W^{(s)} = [W^{(s)}_{ij}]_{i,j=1}^{N}$, where each element $W^{(s)}_{ij}$ depends on the chosen weighting scheme. Some examples of weighting schemes include:
\begin{itemize}
    \item Similarity weights: $W^{(s)}_{ij} = (M^{(s)} - D^{(s)}_{ij}) \mathbb{I} ( G_{ij}^{(s)} \neq 0)$, where $M^{(s)} = \max_{i,j} D^{(s)}_{ij}$ and $\mathbb{I}(\cdot)$ is the indicator function;
    \item Dissimilarity-induced kernel weights: $W^{(s)}_{ij} = \exp(-D^{(s)}_{ij}/\sigma^{(s)}) \mathbb{I} ( G_{ij}^{(s)} \neq 0)$, where $\sigma^{(s)}$ is a bandwidth parameter; 
    \item Graph-induced rank weights \citep{zhou2023new}: for the $k$-NNG, $W^{(s)}_{ij} = (k - l + 1) \mathbb{I} ( G_{ij}^{(s)} \neq 0)$ if the observation $j$ corresponds to the $l$th nearest neighbor of $i$ for some $1 \leq l \leq k$, and $W^{(s)}_{ij} = 0$ otherwise. 
\end{itemize}

After we obtain the multi-view weighted graphs, the proposed method aggregates information from all views based on the following quantities for each view $s = 1,\ldots,S$, where $S$ is the total number of views: 
\begin{equation*}
    U_x^{(s)} = \sum_{i=1}^{m} \sum_{j=1}^{m} W^{(s)}_{ij} 
    \qquad \text{and} \qquad 
    U_y^{(s)} = \sum_{i=m+1}^{N} \sum_{j=m+1}^{N} W^{(s)}_{ij}.
\end{equation*}
The proposed test statistic is defined as $T_S = \bv_S \trans \bSigma_S^{-1} \bv_S$, where
\begin{equation*}
    \bv_S = (
    U_x^{(1)} - \mu_x^{(1)}, U_y^{(1)} - \mu_y^{(1)},
    \ldots, 
    U_x^{(S)} - \mu_x^{(S)}, U_y^{(S)} - \mu_y^{(S)}
    )\trans,
\end{equation*}
with $\mu_x^{(s)} = \EE(U_x^{(s)})$, $\mu_y^{(s)} = \EE(U_y^{(s)})$, and $\bSigma_S = \Cov(\bv_S)$. We use $\EE$, $\Var$, and $\Cov$ to denote the expectation, variance, and covariance under the permutation null distribution, respectively. The closed-form expressions for the means and covariance matrix are presented in \Cref{thm:meanvar} in \Cref{sec:theory}. The Mahalanobis-type test statistic is designed to include deviations of any view in either direction from their expectations under the permutation null, so that a larger test statistic indicates more distributional differences.

Our framework is highly flexible, allowing each view to use different combinations of dissimilarity metrics, graph types, and weighting schemes. For instance, we can apply the same dissimilarity metric for all views but vary the graph structure (e.g., using $k$-NNG versus $k$-MST) or modify the hyperparameters, such as the graph size $k$. Alternatively, we can keep the distance metric and graph structure consistent across views while varying the weighting schemes. Our theoretical analysis in \Cref{sec:theory}  accommodates these variations as long as the weight matrices $\W^{(s)}$ satisfy mild conditions. We do not aim to provide a comprehensive list of view configurations, nor do we prescribe an optimal combination of views. Instead, the proposed test framework allows for adaptable choices that can be customized based on domain knowledge or specific needs of the analysis. The exploration of these configurations and their optimal selection is left as a subject for future work.

\begin{remark}
\label{rem1}
    Although we present the proposed test in the context of high-dimensional Euclidean data and define the views based on the moments of the data, the proposed test is also useful for non-Euclidean data. We give three examples. 
    For distributional data, we can use different orders of  Wasserstein distance as the dissimilarity measures for different views. For image data, the Euclidean distance may be used to measure the pixel-wise difference, while the cosine similarity may be used to measure the angle between the pixel vectors; for colored images, the difference may also be based on each channel of the red-green-blue (RGB) color space. For network data, the cosine similarity and the Euclidean distance of their adjacency matrices can be used as two views to measure the difference between the networks.
\end{remark}

\begin{remark}
    The recent work of \cite{chatterjee2024boosting} proposed the mMMD method to boost the power of kernel-based tests by combining maximum mean discrepancy estimates over multiple kernels using their Mahalanobis distances. While mMMD strengthens traditional single-kernel methods through considering multiple choices of bandwidths, our framework, MATES, takes a fundamentally different perspective by allowing the use of various similarity measures that directly target unique distributional characteristics. For example, MATES can incorporate the Manhattan distance on higher-order moments, making it particularly effective when the distributions differ in these moments. This flexibility enables MATES to achieve significantly higher power than existing methods across diverse scenarios, as demonstrated in \Cref{sec:simulation}. On the contrary, mMMD is motivated by the limitations of the median heuristic \citep{gretton2012optimal,ramdas2015adaptivity}, and it aggregates multiple kernels with both small and large bandwidths. As shown in \Cref{tab:example} and later in \Cref{sec:simulation}, even with multiple choices of bandwidths, the mMMD test lacks power 
    when the two samples differ primarily in their high-order moments.
    
    Moreover, the limiting distribution of mMMD under the null hypothesis depends on the unknown underlying distributions. Consequently, mMMD requires a bootstrap or permutation approach to obtain the $p$-value. In contrast, our test statistic is asymptotically distribution-free (\Cref{thm:limitDist}), enabling straightforward and efficient $p$-value computation. %
 \end{remark}

\section{Performance Analysis}\label{sec:simulation}

In this section, we conduct numerical experiments to evaluate the performance of the proposed method. In the experiments, we choose $S=4$ views, with the $s$th view constructed using the Manhattan distance based on the $s$th moment, as detailed in Section \ref{sec:method}. This choice strikes a balance between capturing diverse aspects of the data and ensuring computational efficiency, as statistical practitioners are often interested in mean, variance, skewness, and kurtosis, while focus less on moments of the the fifth order or higher. The $k$-NNG with $k = \lfloor N^{0.8} \rfloor$ is used for the similarity graph, where $\lfloor x \rfloor$ denotes the largest integer not greater than $x$. We use this choice of $k$ as it leads to good power performance in general. A detailed discussion about the choice of $k$ is presented in Supplementary S.1.1. The graph is weighted by the dissimilarity-induced kernel with the bandwidth chosen to be the median of the pairwise dissimilarities of the pooled sample. This median heuristic is commonly used for various methods \citep[see, for example,][]{gretton2012kernel,song2023generalized,yan2023kernel} and has been shown to yield stable results. In our numerical analysis, we focus on the $k$-NNG, while additional results for $k$-MST and r$k$-NNG are provided in Supplementary S.1.2. 

Comparisons of type-I error and empirical power are made with the $12$  methods mentioned in \Cref{sec:intro}, namely CM, GET, BD, GED, RF, MT, GPK, RISE, MMD, xMMD, aMMD, and mMMD. %
All methods are implemented with their default choices of tuning parameters. We use the asymptotic distribution of
$T_S$ to obtain the $p$-value for MATES, which will be given in Theorem \ref{thm:limitDist}.

We generate $\X = (X_1,\ldots,X_d)$ and $\Y = (Y_1,\ldots,Y_d) \in \RR^{d}$ with $d \in \{200,500,1000\}$, and the sample size is $m = 50$ and $n = 50$ or $100$. The significance level is set to $\alpha = 0.05$, and the number of replications is $1000$. To show MATES can control the type-I error well with the distribution-free limiting distribution, we include several scenarios (a to e) with various types of distributions. Scenarios for the alternative hypothesis include four settings that target different types of distributions (I to IV), and another setting (V) for the same distribution but with different parameters. Various types of signals in distributional differences (i to iv) are also considered to give a comprehensive comparison. We set the parameters of the distributions to make the tests have moderate power to be comparable. These settings are summarized in \Cref{tab:simsetting}.

\begin{table}[!htbp]
    \centering\spacingset{1}
    \caption{Simulation settings.}
    \label{tab:simsetting}
    \resizebox{\textwidth}{!}{%
    \begin{tabular}{ccc}
    \toprule 
    Hypothesis & Setting & Explanation \\\midrule
    \multirow{5}{*}{Null} 
         & (a)  & Normal distributions\\
         & (b)  & Gaussian mixture distributions  \\
         & (c)  & Generalized normal distributions  \\
         & (d)  & $t$-distributions \\
         & (e)  & Gamma distributions  \\
    \midrule
    \multirow{10}{*}{Alternative} & (I)  & Normal versus $t$ distributions \\
         & (II) & Normal versus Gaussian mixture distributions \\
         & (III) & Normal versus generalized normal distributions \\
         & (IV) & Log-normal versus gamma distributions \\
         & (V)  & $t$-distributions with different parameters \\
         & (i)  & Full difference: Distributional differences in all dimensions \\
         & (ii) & Partial difference: Distributional differences in one third of the dimensions \\
         & (iii) & Correlated difference: Dimensions are correlated \\
         & (iv) & Directional difference: Strength of the difference varies across dimensions \\
    \bottomrule
    \end{tabular}%
    }
\end{table}

The data generation scheme for the null hypotheses is as follows:
\begin{enumerate}[label=(\alph*)]
    \item Normal distributions: $\X,\Y \sim N(\mathbf{0}_d, \mathbf{I}_d)$;
    
    \item Gaussian mixture distributions: $\X$ and $\Y$ are independently generated from a mixture of $N(0.5 \mathbf{1}_d, \mathbf{I}_d)$ and $N(- 0.5 \mathbf{1}_d, \mathbf{I}_d)$ with equal probability;  

    \item Generalized normal distributions: for $j=1,\ldots,d$, $X_{j}$ and $Y_{j}$ are independently generated from a generalized normal distribution, which has the density function 
    \begin{align*}
        f(x) = \frac{\beta}{2 \alpha \Gamma(1/\beta)} \exp\left( - \left( \frac{|x-\mu|}{\alpha} \right)^\beta \right),
    \end{align*}
    with parameter $\mu=0$, $\alpha=1$, and $\beta=3$;

    \item $t$-distributions: for $j=1,\ldots,d$, $X_{j}, Y_{j} \sim t_{15}$ independently;
    
    \item Gamma distributions: for $j=1,\ldots,d$, $X_{j}$ and $Y_{j}$ are independently generated from a gamma distribution with shape $2$ and rate $2$.
\end{enumerate}
\Cref{tab:null} shows the empirical sizes of all methods for $m=n=50$. It can be seen that MATES can control the type-I error well under various settings, which shows the validity of the asymptotic approximation even for relatively small sample sizes. Results presented in Supplementary S.1.3 show a similar pattern for MATES with other combinations of sample sizes, similarity graphs, and distance metrics. Other methods also have sizes near the nominal level $\alpha$, except for the mMMD test that is relatively conservative under high dimensional cases.

\begin{table}[!htbp]
    \centering\spacingset{1}
    \caption{Empirical size (in percent) with $\alpha= 0.05$ and $m = n = 50$.}
    \resizebox{\textwidth}{!}{%
        \begin{tabular}{cccccccccccccc}
    \toprule
    $d$  & MATES & CM   & GET  & BD   & GED  & RF   & MT   & GPK  & RISE & MMD  & xMMD & aMMD & mMMD \\
    \midrule
    \multicolumn{14}{c}{Setting (a)} \\
    200  & 5.3  & 3.4  & 5.7  & 5.4  & 5.8  & 6.1  & 5.5  & 5.9  & 5.7  & 5.7  & 5.3  & 5.6  & 2.1 \\
    500  & 5.3  & 3.1  & 4.0  & 5.8  & 4.9  & 4.5  & 4.5  & 4.8  & 3.4  & 3.9  & 5.0  & 3.8  & 1.7 \\
    1000 & 5.0  & 3.4  & 3.7  & 5.4  & 5.4  & 4.8  & 5.2  & 4.3  & 3.7  & 4.4  & 4.4  & 4.8  & 1.4 \\
    \midrule
    \multicolumn{14}{c}{Setting (b)} \\
    200  & 5.5  & 3.5  & 3.5  & 3.9  & 4.6  & 6.4  & 4.3  & 4.0  & 3.4  & 4.7  & 6.6  & 3.8  & 1.7 \\
    500  & 4.5  & 3.4  & 3.8  & 4.6  & 4.4  & 5.2  & 5.6  & 4.4  & 3.7  & 5.2  & 6.5  & 5.5  & 0.9 \\
    1000 & 6.5  & 3.3  & 5.7  & 5.1  & 5.6  & 4.5  & 5.1  & 4.7  & 5.4  & 4.4  & 4.6  & 4.6  & 1.7 \\
    \midrule
    \multicolumn{14}{c}{Setting (c)} \\
    200  & 5.2  & 2.9  & 5.7  & 4.4  & 5.0  & 5.4  & 4.1  & 4.6  & 5.5  & 6.2  & 4.8  & 5.5  & 2.2 \\
    500  & 5.9  & 4.0  & 5.1  & 5.8  & 6.1  & 5.7  & 4.7  & 4.9  & 3.9  & 5.1  & 5.3  & 5.8  & 1.2 \\
    1000 & 5.1  & 4.5  & 4.9  & 4.8  & 4.2  & 5.7  & 4.8  & 4.0  & 5.3  & 4.2  & 5.6  & 4.1  & 1.2 \\
    \midrule
    \multicolumn{14}{c}{Setting (d)} \\
    200  & 6.0  & 3.8  & 4.1  & 5.4  & 4.4  & 5.8  & 4.8  & 4.8  & 4.6  & 4.6  & 4.9  & 5.5  & 1.0 \\
    500  & 5.2  & 3.5  & 4.6  & 4.2  & 5.2  & 6.2  & 5.1  & 4.0  & 4.1  & 5.1  & 6.5  & 5.2  & 2.2 \\
    1000 & 6.6  & 3.0  & 5.0  & 4.3  & 4.5  & 4.8  & 4.2  & 3.8  & 4.2  & 5.0  & 5.8  & 4.9  & 2.0 \\
    \midrule
    \multicolumn{14}{c}{Setting (e)} \\
    200  & 5.7  & 3.8  & 4.8  & 4.3  & 5.8  & 5.1  & 5.9  & 3.5  & 4.1  & 5.6  & 5.7  & 4.3  & 1.8 \\
    500  & 4.5  & 4.8  & 3.7  & 4.9  & 4.7  & 5.9  & 5.2  & 4.7  & 3.9  & 5.9  & 6.0  & 5.7  & 1.3 \\
    1000 & 4.9  & 4.0  & 4.0  & 5.8  & 5.3  & 6.4  & 5.6  & 3.6  & 4.4  & 4.2  & 5.9  & 4.4  & 1.5 \\
    \bottomrule
    \end{tabular}%
    }
    \label{tab:null}%
\end{table}%

The data generation scheme for the alternative hypotheses is as follows:
\begin{enumerate}[label=(\Roman*)]
    \item Normal versus $t$: 
    \begin{enumerate}[label=(\roman*)]
        \item Full difference: for $j=1,\ldots,d$, $X_{j} \sim t_{15}$ ($d=200$), $X_{j} \sim t_{25}$ ($d=500$), or $X_{j} \sim t_{35}$ ($d=1000$), and $Y_{j}$ is normally-distributed with the same mean and variance as $X_{j}$. This is the setting that we considered for the motivating example in \Cref{tab:example}.
        
        \item Partial difference: for $j=1,\ldots,d$, $X_{j} \sim t_{5}$ ($d=200$), $X_{j} \sim t_{7.5}$ ($d=500$), or $X_{j} \sim t_{10}$ ($d=1000$). For $j=1,\ldots,\lceil d/3 \rceil$, $Y_{j}$ is normally-distributed with the same mean and variance as $X_{j}$, and for the remaining $j$, $Y_{j}$ has the same distribution as $X_{j}$;
        
        \item Correlated difference: for $j=1,\ldots,d$, $X_{j} \sim t_{15}$ ($d=200$), $X_{j} \sim t_{25}$ ($d=500$), or $X_{j} \sim t_{35}$ ($d=1000$), $Y'_{j}$ is normally-distributed with the same mean and variance as $X_{j}$, and $\Y = \bSigma^{1/2} \Y'$ with $\bSigma_{ij} = 0.1^{|i-j|}$;
        
        \item Directional difference: for $j=1,\ldots,d$, $X_{j} \sim t_{\df_j}$ with $\df_j \sim \mathrm{Unif}(5,40)$ ($d=200$), $\df_j \sim \mathrm{Unif}(8,60)$ ($d=500$), or $\df_j \sim \mathrm{Unif}(15,80)$ ($d=1000$), and $Y_{j}$ is normally-distributed with the same mean and variance as $X_{j}$.
    \end{enumerate}

    \item Normal versus Gaussian mixture:
    \begin{enumerate}[label=(\roman*)]
        \item Full difference: for $j=1,\ldots,d$, $X_{j}$ is from a mixture of two normal distributions with equal probability of means $\mu_X$ and $-\mu_X$ and variance $1$, where $\mu_X = 0.78$ ($d=200$), $\mu_X = 0.65$ ($d=500$), or $\mu_X = 0.6$ ($d=1000$), and $Y_{j}$ is normally-distributed with the same mean and variance as $X_{j}$;%
        
        \item Partial difference: for $j=1,\ldots,d$, $X_{j}$ is from a mixture of two normal distributions with equal probability of means $\mu_X$ and $-\mu_X$ and variance $1$, where $\mu_X = 1.15$ ($d=200$), $\mu_X = 0.95$ ($d=500$), or $\mu_X = 0.85$ ($d=1000$), for $j=1,\ldots,\lceil d/3 \rceil$, $Y_{j}$ is normally-distributed with the same mean and variance as $X_{j}$, and for the remaining $j$, $Y_{j}$ is iid distributed as $X_{j}$; 
        
        \item Correlated difference: for $j=1,\ldots,d$, $X_{j}$ is from a mixture of two normal distributions with equal probability of means $\mu_X$ and $-\mu_X$ and variance $1$, where $\mu_X = 0.8$ ($d=200$), $\mu_X = 0.66$ ($d=500$), or $\mu_X = 0.6$ ($d=1000$), $Y'_{j}$ is normally-distributed with the same mean and variance as $X_{j}$, and $\Y = \bSigma^{1/2} \Y'$ with $\bSigma_{ij} = 0.1^{|i-j|}$;
        
        \item Directional difference: for $j=1,\ldots,d$, $X_{j}$ is from a mixture of two normal distributions with equal probability of means $\mu_X$ and $-\mu_X$ and variance $1$, where $\mu_X \sim \mathrm{Unif}(0,1.18)$ ($d=200$), $\mu_X \sim \mathrm{Unif}(0,1.0)$ ($d=500$), or $\mu_X \sim \mathrm{Unif}(0,0.9)$ ($d=1000$), and $Y_{j}$ is normally-distributed with the same mean and variance as $X_{j}$.
    \end{enumerate}

    \item Normal versus generalized normal: 
    \begin{enumerate}[label=(\roman*)]
        \item Full difference: for $j=1,\ldots,d$, $X_{j}$ is from a generalized normal distribution with parameter $\mu=0$, $\alpha=1$, and $\beta$ of $2.4$ ($d=200$), $2.2$ ($d=500$), or $2.15$ ($d=1000$), and $Y_{j}$ is normally-distributed with the same mean and variance as $X_{j}$;
        
        \item Partial difference: for $j=1,\ldots,d$, $X_{j}$ is from a generalized normal distribution with parameter $\mu=0$, $\alpha=1$, and $\beta$ of $3.2$ ($d=200$), $2.7$ ($d=500$), or $2.5$ ($d=1000$). For $j=1,\ldots,\lceil d/3 \rceil$, $Y_{j}$ is normally-distributed with the same mean and variance as $X_{j}$, and for the remaining $j$, $Y_{j}$ has the same distribution as $X_{j}$;
        
        \item Correlated difference: for $j=1,\ldots,d$, $X_{j}$ is from a generalized normal distribution with parameter $\mu=0$, $\alpha=1$, and $\beta$ of $2.4$ ($d=200$), $2.2$ ($d=500$), or $2.15$ ($d=1000$), $Y'_{j}$ is normally-distributed with the same mean and variance as $X_{j}$, and $\Y = \bSigma^{1/2} \Y'$ with $\bSigma_{ij} = 0.1^{|i-j|}$; 
        
        \item Directional difference: for $j=1,\ldots,d$, $X_{j}$ is from a generalized normal distribution with parameter $\mu=0$, $\alpha=1$, and $\beta$ from $\mathrm{Unif}(2,2.9)$ ($d=200$), $\mathrm{Unif}(2,2.5)$ ($d=500$), or $\mathrm{Unif}(2,2.3)$ ($d=1000$), and $Y_{j}$ is normally-distributed with the same mean and variance as $X_{j}$.
    \end{enumerate}

    \item Lognormal versus gamma:
    \begin{enumerate}[label=(\roman*)]
        \item Full difference: for $j=1,\ldots,d$, $X_{j}$ is lognormally-distributed with $\mu=0$ and $\sigma=0.24$ ($d=200$), $\sigma=0.19$ ($d=500$), or $\sigma=0.14$ ($d=1000$), and $Y_{j}$ is gamma-distributed with the same mean and variance as $X_{j}$;
        
        \item Partial difference: for $j=1,\ldots,d$, $X_{j}$ is lognormally-distributed with $\mu=0$ and $\sigma=0.5$ ($d=200$), $\sigma=0.42$ ($d=500$), or $\sigma=0.3$ ($d=1000$). For $j=1,\ldots,\lceil d/3 \rceil$, $Y_{j}$ is gamma-distributed with the same mean and variance as $X_{j}$, and for the remaining $j$, $Y_{j}$ has the same distribution as $X_{j}$;
        
        \item Correlated difference: for $j=1,\ldots,d$, $X_{j}$ is lognormally-distributed with $\mu=0$ and $\sigma=0.3$ ($d=200$), $\sigma=0.23$ ($d=500$), or $\sigma=0.19$ ($d=1000$), $Y'_{j}$ is gamma-distributed with the same mean and variance as $X_{j}$, and $\Y = \bSigma^{1/2} \Y'$ with $\bSigma_{ij} = 0.005^{|i-j|}$;
        
        \item Directional difference: for $j=1,\ldots,d$, $X_{j}$ is lognormally-distributed with $\mu=0$ and $\sigma$ from $\mathrm{Unif}(0.01,0.45)$ ($d=200$), $\mathrm{Unif}(0.01,0.32)$ ($d=500$), or $\mathrm{Unif}(0.01,0.22)$ ($d=1000$), and $Y_{j}$ is gamma-distributed with the same mean and variance as $X_{j}$.
    \end{enumerate}

    \item $t$ distributions: for $j=1,\ldots,d$, $X'_{j} \sim t_{5}$ and $X_{j}$ is obtained by standardizing $X'_{j}$ to have mean $0$ and variance $1$, and 
    \begin{enumerate}[label=(\roman*)]
        \item Full difference: for $j=1,\ldots,d$, $Y'_{j} \sim t_{\df_j}$ with $\df_j = 6.8$ ($d=200$), $\df_j = 6.2$ ($d=500$), or $\df_j = 5.8$ ($d=1000$), and $Y_{j}$ is obtained by standardizing $Y'_{j}$ to have mean $0$ and variance $1$;
        
        \item Partial difference: for $j=1,\ldots,\lceil d/3 \rceil$, $Y'_{j} \sim t_{\df_j}$ with $\df_j = 100$ ($d=200$), $\df_j = 13$ ($d=500$), or $\df_j = 9$ ($d=1000$), and for the remaining $j$, $Y'_{j} \sim t_{5}$, and $Y_{j}$ is obtained by standardizing $Y'_{j}$ to have mean $0$ and variance $1$;
        
        \item Correlated difference: for $j=1,\ldots,d$, $Y'_{j} \sim t_{\df_j}$ with $\df_j = 6.8$ ($d=200$), $\df_j = 6.1$ ($d=500$), or $\df_j = 5.8$ ($d=1000$), and $\Y = \bSigma^{1/2} \Y''$ with $\bSigma_{ij} = 0.1^{|i-j|}$, where $\Y''$ is obtained by standardizing $Y'_{j}$ to have mean $0$ and variance $1$;
        
        \item Directional difference: for $j=1,\ldots,d$, $Y'_{j} \sim t_{\df_j}$ with $\df_j \sim \mathrm{Unif}(5,9)$ ($d=200$), $\df_j \sim \mathrm{Unif}(5,7.4)$ ($d=500$), or $\df_j \sim \mathrm{Unif}(5,6.7)$ ($d=1000$), and $Y_{j}$ is obtained by standardizing $Y'_{j}$ to have mean $0$ and variance $1$.
    \end{enumerate}

\end{enumerate}

The empirical power of all methods  for $m=n=50$ is presented in \Cref{tab:patterni,tab:patternii,tab:patterniii,tab:patterniv}. Results for $m=50$ and $n=100$ show similar trends, and are presented in Supplementary S.1.4. We highlight the highest power for each setting in boldface.  

The proposed method outperforms the other methods in all the settings considered, and it has stable performances across different dimensions and difference patterns. Tests such as BD, GPK, MMD, and xMMD have power lower than 0.1 in most settings, indicating that their reliance on the common choice of Euclidean distance metric %
fails to detect differences in higher moments in the distributions. The GED test has relatively high power in some settings. For example, 
for (ii)(I)(II)(IV)(V), (iii)(IV)(V), and (iv)(IV), the GED test has moderate power (from 0.2 to 0.5), and performs better than the other tests. However, in general, the power of GED decreases when the dimension increases, which supposedly should provide more information for distinguishing distributional differences; in contrast, the proposed method has comparable power and exhibits stability with higher dimensions. Tests like GET, RISE, and aMMD are only slightly powerful under some alternative hypothesis, and show a decreasing trend as the dimension increases, which is similar to GED.

We also observe that more complicated methods such as the random forest cannot capture differences in the overall distribution as well. Specifically, the results of RF show that the random forest-based method still has relatively low power in most settings. It has slightly higher power in (ii), the ``partial'' difference pattern, with distribution types (I)(IV)(V), but similar to the GED method, the higher the dimension is, the less powerful it becomes. On the other hand, existing deep learning-based two-sample testing methods are not directly applicable to the current setting. For example, 
\cite{kirchler2020two} requires a pre-trained deep learning model, which is not readily available for the type of data we considered in the simulation.

\begin{table}[!htbp]
    \centering\spacingset{1}
    \caption{Empirical power (in percent) with $\alpha= 0.05$ and  $m = n = 50$ under the difference pattern (i) full difference. }
    \resizebox{\textwidth}{!}{%
    \begin{tabular}{cccccccccccccc}
    \toprule
    $d$  & MATES & CM   & GET  & BD   & GED  & RF   & MT   & GPK  & RISE & MMD  & xMMD & aMMD & mMMD \\
    \midrule
    \multicolumn{14}{c}{Setting (I)} \\
    200  & \textbf{90.9} & 4.5  & 6.3  & 5.2  & 9.3  & 6.6  & 5.5  & 4.1  & 6.1  & 4.6  & 5.5  & 5.5  & 1.3 \\
    500  & \textbf{85.2} & 3.9  & 4.4  & 5.6  & 7.1  & 6.1  & 4.5  & 4.8  & 4.5  & 6.2  & 6.0  & 6.1  & 1.3 \\
    1000 & \textbf{85.8} & 3.3  & 4.7  & 5.5  & 6.8  & 6.6  & 5.7  & 4.8  & 4.3  & 5.2  & 6.3  & 5.7  & 0.9 \\
    \midrule
    \multicolumn{14}{c}{Setting (II)} \\
    200  & \textbf{81.2} & 3.4  & 5.2  & 3.6  & 9.8  & 5.7  & 5.2  & 4.1  & 5.2  & 5.5  & 5.2  & 6.0  & 2.1 \\
    500  & \textbf{71.3} & 2.9  & 5.2  & 5.6  & 8.0  & 5.4  & 5.4  & 4.5  & 4.1  & 4.9  & 6.3  & 6.1  & 1.5 \\
    1000 & \textbf{82.4} & 2.9  & 5.3  & 5.4  & 7.0  & 4.8  & 5.3  & 4.8  & 5.7  & 4.5  & 5.4  & 5.1  & 1.6 \\
    \midrule
    \multicolumn{14}{c}{Setting (III)} \\
    200  & \textbf{89.5} & 2.6  & 4.7  & 4.6  & 10.4 & 6.8  & 5.5  & 3.9  & 5.1  & 4.4  & 5.7  & 5.3  & 1.8 \\
    500  & \textbf{82.5} & 3.0  & 5.4  & 4.5  & 8.9  & 7.1  & 4.5  & 4.1  & 6.3  & 5.8  & 5.7  & 5.7  & 2.0 \\
    1000 & \textbf{82.7} & 3.2  & 4.8  & 5.2  & 5.4  & 5.7  & 5.5  & 4.1  & 4.4  & 4.6  & 6.3  & 4.1  & 1.2 \\
    \midrule
    \multicolumn{14}{c}{Setting (IV)} \\
    200  & \textbf{74.3} & 4.6  & 6.0  & 6.0  & 14.0 & 7.6  & 5.4  & 4.6  & 6.2  & 4.6  & 5.9  & 6.6  & 1.2 \\
    500  & \textbf{89.9} & 3.8  & 5.6  & 6.6  & 12.6 & 7.5  & 3.9  & 6.0  & 5.7  & 5.1  & 5.1  & 6.8  & 1.5 \\
    1000 & \textbf{84.7} & 3.8  & 4.9  & 4.8  & 10.3 & 6.9  & 3.9  & 4.8  & 5.1  & 5.0  & 6.1  & 6.1  & 2.4 \\
    \midrule
    \multicolumn{14}{c}{Setting (V)} \\
    200  & \textbf{77.9} & 5.1  & 10.0 & 6.2  & 13.4 & 7.5  & 5.0  & 4.3  & 9.6  & 5.4  & 6.4  & 7.4  & 1.3 \\
    500  & \textbf{88.3} & 4.2  & 7.6  & 5.1  & 11.7 & 5.7  & 4.5  & 4.3  & 8.6  & 4.6  & 5.1  & 6.4  & 0.9 \\
    1000 & \textbf{87.8} & 3.9  & 6.1  & 4.3  & 9.8  & 5.2  & 4.5  & 3.5  & 6.4  & 5.5  & 6.9  & 6.5  & 0.9 \\
    \bottomrule
    \end{tabular}%
    }
    \label{tab:patterni}%
\end{table}%

\begin{table}[!htbp]
    \centering\spacingset{1}
    \caption{Empirical power (in percent) with $\alpha= 0.05$ and  $m = n = 50$ under the difference pattern (ii) partial difference.}
    \resizebox{\textwidth}{!}{%
    \begin{tabular}{cccccccccccccc}
    \toprule
    $d$  & MATES & CM   & GET  & BD   & GED  & RF   & MT   & GPK  & RISE & MMD  & xMMD & aMMD & mMMD \\
    \midrule
    \multicolumn{14}{c}{Setting (I)} \\
    200  & \textbf{80.3} & 2.9  & 7.4  & 4.8  & 43.4 & 14.3 & 4.3  & 3.4  & 7.4  & 5.4  & 6.2  & 12.6 & 0.7 \\
    500  & \textbf{78.4} & 3.3  & 5.9  & 5.4  & 19.0 & 9.7  & 5.5  & 4.2  & 6.1  & 4.5  & 5.5  & 7.7  & 1.2 \\
    1000 & \textbf{84.4} & 2.9  & 5.0  & 3.8  & 16.2 & 7.3  & 4.9  & 3.4  & 4.6  & 4.1  & 5.0  & 7.0  & 1.3 \\
    \midrule
    \multicolumn{14}{c}{Setting (II)} \\
    200  & \textbf{83.7} & 3.3  & 4.4  & 3.9  & 20.6 & 9.3  & 4.9  & 4.1  & 4.2  & 5.2  & 5.5  & 7.2  & 1.4 \\
    500  & \textbf{79.0} & 5.0  & 4.2  & 5.1  & 12.6 & 5.5  & 4.3  & 2.9  & 4.0  & 4.3  & 5.1  & 4.9  & 0.7 \\
    1000 & \textbf{82.2} & 4.7  & 5.5  & 4.9  & 9.8  & 5.6  & 4.6  & 4.6  & 5.2  & 4.5  & 5.0  & 6.3  & 1.1 \\
    \midrule
    \multicolumn{14}{c}{Setting (III)} \\
    200  & \textbf{83.3} & 2.8  & 4.6  & 4.1  & 15.2 & 8.4  & 4.5  & 4.0  & 4.4  & 5.4  & 4.9  & 6.4  & 1.9 \\
    500  & \textbf{82.2} & 3.7  & 5.4  & 5.1  & 11.6 & 7.3  & 4.4  & 4.3  & 5.1  & 4.6  & 6.8  & 6.2  & 1.4 \\
    1000 & \textbf{89.8} & 4.5  & 5.5  & 5.7  & 9.6  & 6.6  & 5.1  & 4.8  & 5.0  & 4.7  & 6.5  & 4.9  & 1.6 \\
    \midrule
    \multicolumn{14}{c}{Setting (IV)} \\
    200  & \textbf{61.3} & 4.1  & 5.7  & 6.5  & 29.6 & 16.1 & 5.6  & 5.2  & 6.5  & 5.0  & 5.6  & 10.7 & 1.9 \\
    500  & \textbf{81.3} & 3.0  & 4.7  & 4.5  & 27.3 & 12.1 & 5.9  & 4.9  & 4.2  & 4.5  & 5.2  & 9.0  & 1.3 \\
    1000 & \textbf{72.3} & 4.7  & 4.7  & 5.9  & 17.7 & 8.9  & 4.6  & 5.0  & 4.8  & 5.3  & 5.9  & 8.4  & 0.9 \\
    \midrule
    \multicolumn{14}{c}{Setting (V)} \\
    200  & \textbf{78.9} & 4.3  & 7.6  & 5.6  & 38.3 & 13.6 & 5.3  & 4.9  & 6.5  & 4.8  & 6.3  & 11.0 & 1.8 \\
    500  & \textbf{80.7} & 4.1  & 5.2  & 5.2  & 26.4 & 10.1 & 5.4  & 3.5  & 5.3  & 4.7  & 5.0  & 9.0  & 1.1 \\
    1000 & \textbf{82.3} & 3.4  & 5.0  & 3.9  & 22.9 & 8.9  & 3.5  & 3.9  & 5.4  & 4.5  & 5.6  & 8.5  & 1.0 \\
    \bottomrule
    \end{tabular}%
    }
    \label{tab:patternii}%
\end{table}%

\begin{table}[!htbp]
    \centering\spacingset{1}
    \caption{Empirical power (in percent) with $\alpha= 0.05$ and  $m = n = 50$ under the difference pattern (iii) correlated difference. }
    \resizebox{\textwidth}{!}{%
    \begin{tabular}{cccccccccccccc}
    \toprule
    $d$  & MATES & CM   & GET  & BD   & GED  & RF   & MT   & GPK  & RISE & MMD  & xMMD & aMMD & mMMD \\
    \midrule
    \multicolumn{14}{c}{Setting (I)} \\
    200  & \textbf{90.8} & 4.9  & 5.6  & 4.6  & 9.2  & 7.3  & 5.4  & 3.6  & 5.3  & 4.5  & 5.7  & 5.9  & 1.6 \\
    500  & \textbf{86.0} & 5.1  & 4.9  & 6.0  & 7.5  & 6.1  & 5.1  & 4.8  & 4.9  & 6.2  & 6.3  & 6.3  & 1.6 \\
    1000 & \textbf{85.8} & 4.5  & 3.8  & 5.3  & 7.2  & 5.9  & 5.2  & 4.8  & 4.4  & 5.6  & 6.1  & 5.8  & 0.9 \\
    \midrule
    \multicolumn{14}{c}{Setting (II)} \\
    200  & \textbf{90.0} & 4.5  & 5.3  & 3.7  & 10.2 & 7.1  & 4.2  & 4.1  & 6.0  & 6.0  & 4.9  & 6.9  & 2.0 \\
    500  & \textbf{75.1} & 4.2  & 5.4  & 5.3  & 8.3  & 7.0  & 5.5  & 4.6  & 4.9  & 5.3  & 6.0  & 6.1  & 1.4 \\
    1000 & \textbf{81.6} & 4.0  & 5.5  & 5.6  & 6.8  & 6.6  & 5.7  & 4.9  & 4.7  & 4.9  & 5.3  & 5.0  & 1.5 \\
    \midrule
    \multicolumn{14}{c}{Setting (III)} \\
    200  & \textbf{91.6} & 3.2  & 5.3  & 5.0  & 10.9 & 7.0  & 6.0  & 4.4  & 5.3  & 4.2  & 5.7  & 5.3  & 1.7 \\
    500  & \textbf{74.2} & 1.9  & 6.4  & 4.9  & 7.7  & 6.8  & 4.4  & 4.3  & 7.1  & 5.5  & 5.4  & 6.0  & 2.1 \\
    1000 & \textbf{81.5} & 4.1  & 4.0  & 5.0  & 6.4  & 5.9  & 5.0  & 4.1  & 4.8  & 4.8  & 6.3  & 4.5  & 1.2 \\
    \midrule
    \multicolumn{14}{c}{Setting (IV)} \\
    200  & \textbf{90.6} & 4.4  & 7.6  & 7.1  & 30.2 & 11.2 & 9.6  & 4.7  & 6.5  & 6.1  & 7.2  & 13.7 & 0.7 \\
    500  & \textbf{88.1} & 5.2  & 5.5  & 5.6  & 32.5 & 11.7 & 19.7 & 5.3  & 5.7  & 6.6  & 6.5  & 14.7 & 1.3 \\
    1000 & \textbf{82.2} & 4.4  & 6.2  & 5.8  & 43.2 & 14.2 & 31.4 & 7.7  & 5.9  & 10.4 & 10.7 & 24.4 & 2.5 \\
    \midrule
    \multicolumn{14}{c}{Setting (V)} \\
    200  & \textbf{80.5} & 4.3  & 9.8  & 6.3  & 15.7 & 8.4  & 4.0  & 4.6  & 9.2  & 5.6  & 5.9  & 7.4  & 1.4 \\
    500  & \textbf{87.6} & 3.6  & 6.7  & 6.2  & 10.4 & 6.7  & 5.0  & 4.5  & 7.4  & 4.3  & 6.4  & 6.0  & 1.3 \\
    1000 & \textbf{90.5} & 4.6  & 4.6  & 3.8  & 10.6 & 5.7  & 4.8  & 3.4  & 5.7  & 6.0  & 6.8  & 6.3  & 1.0 \\
    \bottomrule
    \end{tabular}%
    }
    \label{tab:patterniii}%
\end{table}%

\begin{table}[!htbp]
    \centering\spacingset{1}
    \caption{Empirical power (in percent) with $\alpha= 0.05$ and  $m = n = 50$ under the difference pattern (iv) directional difference.}
    \resizebox{\textwidth}{!}{%
    \begin{tabular}{cccccccccccccc}
    \toprule
    $d$  & MATES & CM   & GET  & BD   & GED  & RF   & MT   & GPK  & RISE & MMD  & xMMD & aMMD & mMMD \\
    \midrule
    \multicolumn{14}{c}{Setting (I)} \\
    200  & \textbf{87.9} & 3.7  & 7.4  & 6.1  & 12.3 & 7.2  & 4.6  & 5.2  & 6.6  & 5.6  & 6.3  & 7.0  & 1.8 \\
    500  & \textbf{85.5} & 3.3  & 6.0  & 5.1  & 7.9  & 5.9  & 5.3  & 4.5  & 6.3  & 5.5  & 5.4  & 5.9  & 1.9 \\
    1000 & \textbf{80.9} & 4.2  & 4.0  & 4.5  & 6.2  & 6.6  & 5.5  & 3.3  & 4.5  & 4.9  & 4.7  & 5.1  & 1.7 \\
    \midrule
    \multicolumn{14}{c}{Setting (II)} \\
    200  & \textbf{84.3} & 3.6  & 5.5  & 4.0  & 11.6 & 7.1  & 4.5  & 5.3  & 4.7  & 6.4  & 5.3  & 6.8  & 2.1 \\
    500  & \textbf{79.0} & 3.6  & 5.3  & 5.8  & 8.4  & 5.7  & 3.8  & 4.5  & 4.8  & 5.0  & 5.5  & 4.9  & 1.2 \\
    1000 & \textbf{81.8} & 4.6  & 5.3  & 6.0  & 7.7  & 6.3  & 3.8  & 5.0  & 4.6  & 4.9  & 5.6  & 4.9  & 1.3 \\
    \midrule
    \multicolumn{14}{c}{Setting (III)} \\
    200  & \textbf{87.8} & 4.3  & 5.1  & 4.8  & 14.3 & 6.6  & 5.6  & 3.9  & 5.0  & 4.9  & 5.8  & 6.7  & 1.5 \\
    500  & \textbf{85.4} & 3.4  & 4.7  & 5.0  & 9.5  & 5.7  & 5.1  & 4.6  & 4.8  & 4.7  & 5.6  & 4.8  & 1.1 \\
    1000 & \textbf{76.5} & 2.9  & 4.8  & 5.0  & 7.5  & 6.3  & 5.1  & 5.3  & 4.5  & 5.8  & 6.6  & 7.3  & 1.4 \\
    \midrule
    \multicolumn{14}{c}{Setting (IV)} \\
    200  & \textbf{82.5} & 4.1  & 12.5 & 7.2  & 18.2 & 10.2 & 6.5  & 4.8  & 11.5 & 4.9  & 5.4  & 10.8 & 1.3 \\
    500  & \textbf{82.5} & 3.5  & 6.3  & 6.0  & 13.4 & 8.0  & 4.6  & 4.4  & 6.2  & 5.7  & 6.0  & 7.5  & 1.2 \\
    1000 & \textbf{70.4} & 2.4  & 5.7  & 4.6  & 10.6 & 7.0  & 4.2  & 3.6  & 5.6  & 4.7  & 5.7  & 6.2  & 1.5 \\
    \midrule
    \multicolumn{14}{c}{Setting (V)} \\
    200  & \textbf{76.9} & 4.3  & 9.2  & 5.8  & 12.8 & 8.0  & 5.2  & 3.6  & 9.4  & 4.8  & 6.5  & 7.0  & 1.2 \\
    500  & \textbf{82.2} & 3.8  & 9.3  & 5.8  & 12.4 & 7.6  & 6.8  & 4.1  & 8.2  & 5.0  & 6.4  & 7.2  & 1.3 \\
    1000 & \textbf{86.6} & 3.9  & 5.3  & 4.8  & 9.6  & 7.4  & 5.5  & 3.1  & 4.7  & 4.7  & 5.8  & 5.3  & 1.5 \\
    \bottomrule
    \end{tabular}%
    }
    \label{tab:patterniv}%
\end{table}%

\section{Theoretical Properties}\label{sec:theory}
In this section, we first give the closed-form expression for various quantities used to construct the test statistic, and specify the conditions for the test statistic to be well-defined. We then show that $T_S$ is asymptotically distribution-free under the null hypothesis, which enables an analytic approximation of the $p$-value and straightforward type-I error control. 

To give the closed-form expression of the mean and variance of the test statistic under the permutation null distribution, we define the following quantities for $s,s' = 1,\ldots,S$:
\begin{equation*}
    W_1^{(s)} = \sum_{i=1}^{N}\sum_{j=1}^{N} W_{ij}^{(s)},
    \qquad 
    W_2^{(s,s')} = \sum_{i=1}^{N}\sum_{j=1}^{N} W_{ij}^{(s)} W_{ij}^{(s')},
    \qquad 
    W_3^{(s,s')} = \sum_{i=1}^{N}\sum_{j=1}^{N}\sum_{r=1}^{N} W_{ij}^{(s)} W_{ir}^{(s')},
\end{equation*}
\begin{equation*}
    \tW_2^{(s,s')} = W_2^{(s,s')} - \frac{W_1^{(s)}W_1^{(s')}}{N(N-1)},   
    \qquad  \text{ and }    \qquad  
    \tW_3^{(s,s')} = W_3^{(s,s')} - \frac{W_1^{(s)}W_1^{(s')}}{N}. 
\end{equation*}
We assume that $\W^{(s)}$ is symmetric; if it is not, we can replace it with $\frac{1}{2}\big(\W^{(s)} + (\W^{(s)})\trans\big)$. This symmetrization does not affect the values of $U_x^{(s)}$ and $U_y^{(s)}$, but it simplifies the calculations of their expectations and covariances.
\begin{theorem}\label{thm:meanvar}
    Under the permutation null distribution, we have 
    \begin{equation*}
        \mu_x^{(s)} = \frac{m(m-1)}{N(N-1)} W_1^{(s)},
        \qquad 
        \mu_y^{(s)} = \frac{n(n-1)}{N(N-1)} W_1^{(s)},
    \end{equation*}
    and the elements of the covariance matrix $\bSigma_S$ are given by 
    \begin{equation*}
        \Cov(U_x^{(s)}, U_x^{(s')}) = C (m-1)
        \left(
        (n-1)  \tW_2^{(s,s')}
        + 2(m-2) \tW_3^{(s,s')}  
        \right),
    \end{equation*}
    \begin{equation*}
        \Cov(U_y^{(s)}, U_y^{(s')}) = C (n-1)
        \left(
        (m-1)  \tW_2^{(s,s')} 
        + 2(n-2) \tW_3^{(s,s')} 
        \right),
    \end{equation*}
    and
    \begin{equation*}
        \Cov(U_x^{(s)}, U_y^{(s')}) = C (m-1) (n-1)
        \left(
        \tW_2^{(s,s')}
        - 2 \tW_3^{(s,s')} 
        \right),
    \end{equation*}
    with $C = \frac{2mn}{N(N-1)(N-2)(N-3)}$, for $s,s' = 1,\ldots,S$. 
\end{theorem}
The proof of Theorem \ref{thm:meanvar} is provided in Supplementary S.2.1. %
We next establish the necessary and sufficient conditions for the test statistic $T_S$ to be well-defined. The key idea is to decompose $T_S$ into two orthogonal components and identify the conditions under which their corresponding covariances are invertible. We first define $
        \bv_{w} = (U_{w}^{(1)} - \EE(U_{w}^{(1)}), \ldots, U_{w}^{(S)} - \EE(U_{w}^{(S)}))\trans$   and   $
        \bv_{\diff} = (U_{\diff}^{(1)} - \EE(U_{\diff}^{(1)}), \ldots, U_{\diff}^{(S)} - \EE(U_{\diff}^{(S)}))\trans
    $, where      \begin{equation*}
        U_{w}^{(s)} = \frac{n-1}{N-2} U_x^{(s)} + \frac{m-1}{N-2} U_y^{(s)}, 
        \qquad \text{and} \qquad 
        U_{\diff}^{(s)} = U_x^{(s)} - U_y^{(s)}.
    \end{equation*}
\begin{theorem}\label{thm:decomposition}
    When $T_S$ is well-defined, we have
    \begin{equation*}
        T_S = \bv_{w}\trans \bSigma_{w}^{-1} \bv_{w} + \bv_{\diff}\trans \bSigma_{\diff}^{-1} \bv_{\diff}, \quad \text{ and } \quad \Cov(\bv_{w}, \bv_{\diff}) = \mathbf{0}, 
    \end{equation*}
where 
    $\bSigma_{w} = \Cov(\bv_{w})$, $\bSigma_{\diff} = \Cov(\bv_{\diff})$, with the $(s,s')$th element of $\bSigma_{w}$ and $\bSigma_{\diff}$ given by
    \begin{equation*}
        (\bSigma_{w})_{ss'} = \frac{2mn(m-1)(n-1)}{N(N-1)(N-2)(N-3)} \left( 
            \tW_2^{(s,s')} 
            - \frac{2}{N-2} \tW_3^{(s,s')}\right),
    \end{equation*}
    and 
    \begin{equation*}
        (\bSigma_{\diff})_{ss'} = \frac{4mn}{N(N-1)} \tW_3^{(s,s')}.
    \end{equation*}
\end{theorem}
The proof of \Cref{thm:decomposition} is given in Supplementary S.2.2. As a special case, when $S = 1$, $T_S$ reduces to $T_S = (Z_w^{(1)})^2 + (Z_{\diff}^{(1)})^2$, where $Z_w^{(1)} = \big( U_{w}^{(1)} - \EE(U_{w}^{(1)}) \big)/\sqrt{\Var(U_{w}^{(1)})}$ and $Z_{\diff}^{(1)} = \big( U_{\diff}^{(1)} - \EE(U_{\diff}^{(1)}) \big)/\sqrt{\Var(U_{\diff}^{(1)})}$, and $\Cov(Z_w^{(1)}, Z_{\diff}^{(1)}) = 0$. Similar decompositions for the Mahalanobis-type test statistics can also be found in \cite{chu2019asymptotic,song2023generalized,zhou2023new}.

\begin{remark}
    With this decomposition, we offer another perspective of why some previous works that typically rely on the $\ell_2$ distance would have limited power under specific alternative scenarios, as illustrated in \Cref{sec:simulation}. 
    As discussed in these works, two primary cases can arise under the alternative hypothesis: (i) both $U_x^{(1)}$ and $U_y^{(1)}$ exceed their null expectations, which is common under location alternatives. In this case, $\sqrt{2} \max \{\sigma_1,\sigma_2\} < \sqrt{\sigma_1^2 + \sigma_2^2 + \upsilon^2}$ (refer to their definitions following \Cref{tab:example}), and $Z_{w}^{(1)}$ will be large; (ii) one statistic exceeds its expectation under null, and the other falls below, which is common with scale alternatives. In this case, $\sqrt{2} \max \{\sigma_1,\sigma_2\} > \sqrt{\sigma_1^2 + \sigma_2^2 + \upsilon^2}$, and $|Z_{\diff}^{(1)}|$ will be large. However, when the two distributions differ primarily in higher-order moments, neither of these two cases would occur, thus leading to low power for these $\ell_2$-based methods. However, we also note here that simply changing the distance metric is insufficient, as the GED test can still lack power as indicated in \Cref{sec:simulation}.
\end{remark}

According to \Cref{thm:decomposition}, for $T_S$ to be well-defined, we must have ${\rm det}(\bSigma_{w}) > 0$ and ${\rm det}(\bSigma_{\diff}) > 0$. Below, we establish the necessary and sufficient conditions for the covariance matrices $\bSigma_{w}$ and $\bSigma_{\diff}$ to be positive-definite. Define 
$$
\begin{aligned}
& W_{i\cdot}^{(s)} = \sum_{j=1}^N W_{ij}^{(s)}, \quad \widetilde W_{i\cdot}^{(s)} = W_{i\cdot}^{(s)} - \frac{W_1^{(s)}}{N},  \quad \W_{v}^{(s)} = (\widetilde W_{1\cdot}^{(s)},\ldots, \widetilde W_{N\cdot}^{(s)})\trans, \text{ and }\\
& \widehat \W^{(s)} = [\widehat W_{ij}^{(s)}]_{i,j=1}^N  \text{ with } \widehat W_{ij}^{(s)} = W_{ij}^{(s)} - \frac{W_{1}^{(s)} }{N(N-1)} -  \frac{\widetilde W_{i\cdot}^{(s)} + \widetilde W_{j\cdot}^{(s)} }{N-2}    \text{ for } i \neq j \text{ and } \widehat W_{ii}^{(s)} = 0. 
\end{aligned}
$$

\begin{theorem}
When $N \geq \max\{3,S\}$, the covariance matrix $\bSigma_{w}$ is positive-definite if and only if the $S$ matrices $\widehat \W^{(1)}$, $\ldots$, $\widehat \W^{(S)}$ are linearly independent, and the covariance matrix $\bSigma_{\diff}$ is positive-definite if and only if the $S$ vectors $\W_{v}^{(1)}$, $\ldots$, $\W_{v}^{(S)}$ are linearly independent. The test statistic $T_S$ is well-defined if and only if both $\bSigma_{w}$ and $\bSigma_{\diff}$ are positive-definite. 
\label{thm:positive}
\end{theorem}

The proof of \Cref{thm:positive} is given in Supplementary S.2.3. The conditions for positive definiteness of $\bSigma_{w}$ and $\bSigma_{\diff}$ have an intuitive interpretation: each view must provide unique information not captured by the others. If some $\widehat \W^{(s)}$ is a linear combination of other $\widehat \W^{(s')}$’s (or similarly for $\W_{v}^{(s)}$ vectors), the $s$th view is redundant and thus the covariance matrix will become singular. In practice, checking these linear independence conditions is straightforward. If the weighted similarity graphs for different views convey unique information, these conditions are typically satisfied. Otherwise, removing redundant views that contribute to covariance matrix singularity is advisable. 
Additionally, from the proof of \Cref{thm:positive}, we find that $\Cov(U_{w}^{(s)}) \propto   \|\widehat \W^{(s)}\|_{\rm F}^2$ and $\Cov(U_{\diff}^{(s)}) \propto \|\W_{v}^{(s)}\|_2$. Thus, $\Cov(U_{w}^{(s)}) = 0$ if $\widehat W_{ij}^{(s)} = 0$ for all $i$ and $j$, and $\Cov(U_{\diff}^{(s)}) = 0$ when all $W_{i\cdot}^{(s)},i=1,\ldots,N$ values are equal.

Before presenting the main theorem, we introduce some additional notation. Let $a_N \precsim b_N$ denote that $a_N \leq C b_N$ for some constant $C$ for all sufficiently large $N$, $a_N/b_N \to 0$ if $\lim_{N \rightarrow \infty} a_N/b_N = 0$, $a_N = O(b_N)$ if $a_N \precsim b_N$ and $b_N \precsim a_N$. 
\begin{theorem}\label{thm:limitDist}
    Let $\W^{(s)}$ %
    be the weighted graph of the $s$th view. When $m, n \to \infty$ with $m/(m+n) \to p \in (0, 1)$, assume the following conditions:
        \begin{enumerate}[label=(\arabic*),ref=(\arabic*)]
        \item \label{cond:2} $\sum_{i=1}^{N} \left( \sum_{j=1}^{N} W_{ij}^{(s)2} \right)^2 \precsim W_2^{(s,s)2}/N$ for each $s$; 
        \item \label{cond:3} $\sum_{i=1}^{N} \big| \widetilde W_{i\cdot}^{(s)} \big|^3 / \tW_3^{(s,s)3/2} \to 0$ for each $s$; 
        \item \label{cond:4} $\max_{i} W_{i\cdot}^{(s)} / \sqrt{W_2^{(s,s)}} \to 0$ for each $s$; 
        \item \label{cond:5} $\sum_{l=1}^{N} \sum_{i\neq j} W_{il} W_{jl} 
        \left( \sum_{s=1}^{S} \frac{| \widetilde W_{i\cdot}^{(s)}|}{\sqrt{\tW_3^{(s,s)}}} \right) 
        \left( \sum_{s=1}^{S} \frac{|\widetilde W_{j\cdot}^{(s)}|}{\sqrt{\tW_3^{(s,s)}}} \right) \to 0$; 
        \item \label{cond:6} $\sum_{i=1}^{N} \sum_{j=1}^{N} \sum_{r\neq i,j} \sum_{l\neq i,j} W_{ji} W_{ir} W_{jl} W_{rl} \to 0$; 
    \end{enumerate}
    where $W_{ij}=\sum_{s=1}^{S} b_0^{(s)} W_{ij}^{(s)}$ with $b_0^{(s)} = O(1/\sqrt{W_2^{(s,s)}})$. The exact definition of $b_0^{(s)}$ is presented in Supplementary S.2.4. We then have $T_S \to \chi^2_{2S}$ in distribution if all the conditions are satisfied.
\end{theorem}
The proof of \Cref{thm:limitDist} is provided in Supplementary S.2.4. Conditions (1) to (3) represent standard assumptions for graph-based two-sample tests when a single view is employed; see \cite{zhou2023new} and \cite{zhu2024limiting} for details. Conditions (4) and (5), which generalize the corresponding conditions of single-view versions, are required due to potential interactions between different views. Basically, these conditions require that there are not too many hubs (i.e., nodes with a large degree) in the graph.

As demonstrated in the following lemma, further simplifications can be achieved for common graph choices. 
\begin{lemma}\label{thm:lemma} 
    Suppose we use the $k$-NNG or $k$-MST with $k = O(N^\beta)$ and $\beta < 1$. If the graph is weighted by distance/kernel with weight of order $O(1)$ or by graph-induced rank, then the conditions in \Cref{thm:limitDist} can be guaranteed by the following sufficient conditions if the maximum degree of the $S$ graphs is bounded by $C k$:
    \begin{enumerate}[label=(\arabic*'),ref=(\arabic*')]
        \item  \label{cond:1'}
        {$\max_{i} \big| \tW_{i\cdot}^{(s)} \big| 
        / \sqrt{\tW_3^{(s,s)}} = O(N^{-\gamma})$} with $\gamma > \beta/2$, for each $s$;

        \item  \label{cond:2'}
        $N_{\rm sq} = o(N^{2\beta + 2})$, where $N_{\rm sq}$ denotes the number of squares in the aggregated graph $\W = [W_{ij}]_{i,j=1}^{N}$.
    \end{enumerate}
\end{lemma}
Since %
{$\tW_3^{(s,s)} = \sum_{i=1}^{N} \tW_{i\cdot}^{(s)2}$}, Condition \Cref{cond:1'} requires that there is no dominating vertex $i$ such that %
{$\tW_{i\cdot}^{(s)2} \approx \tW_3^{(s,s)}$}. As a special case, if all %
{$\tW_{i\cdot}^{(s)}$}'s are of the same order, then $\gamma = 1/2$, and Condition \Cref{cond:1'} is satisfied automatically. %
On the other hand, for an $S$-view aggregated graph with $N$ vertices and $O(SNk)$ edges, there are at most $O(S^2 N^2 k^2)$ squares. \Cref{cond:2'} requires the absence of this extreme case. The proof of \Cref{thm:lemma} is provided in Supplementary S.2.5.

\section{Analysis of S\&P100 Data}\label{sec:realdata}
When analyzing historical stock returns, investors typically focus on specific time periods to identify patterns, assess risks, and uncover investment opportunities. While a positive mean excess return, or ``alpha'', is desirable in portfolio management, it is usually difficult to achieve in efficient markets as prices quickly adjust to market news, leaving little room for arbitrage. On the other hand, variance is typically well-controlled through diversification, which helps minimize idiosyncratic risks. However, returns may exhibit differences in higher-order moments, such as skewness and kurtosis. Skewness reflects the asymmetry of the return distribution, offering insights into potential risks and rewards beyond mean and variance. Kurtosis, which measures the tailedness or peakedness of the distribution, indicates the likelihood of extreme gains and losses. A higher kurtosis suggests a greater chance of extreme events, while a low value is typical with more stable stocks. As a result, investors often favor positive skewness and avoid high kurtosis. Understanding these higher-order differences is crucial for measuring extreme risks and tail dependencies, based on which investors can have a comprehensive understanding of the potential return and make more informed decisions.

In this section, we use an example to illustrate the importance of incorporating multiple views for detecting differences in stock trends, using the daily closing prices of companies in the S\&P100 obtained from Yahoo Finance\footnote{\href{https://www.kaggle.com/datasets/alessandrolobello/all-s-and-p100-open-price-stocks-forecast}{Open Price Stocks - All S\&P100 trends}} collected before and after the release of ChatGPT on November 30, 2022\footnote{\href{https://openai.com/index/chatgpt/}{Introducing ChatGPT}}.
Studies have shown that large language models (LLMs) like ChatGPT are able to capture subtle sentiment from news headlines and other textual data, outperforming traditional methods in predicting stock returns \citep{chen2023chatgpt,wang2024modeling,lefort2024can}. The stock return patterns may change after ChatGPT's release as its capabilities in analyzing and predicting stock movements may reshape investor behaviors and market dynamics.
In the meantime, the increasing accessibility of technologies such as LLMs may shift investors' focus towards the related industries, especially those sectors connected to AI advancements and data-driven analysis, which may have the potential to cause fluctuations in stock prices. 
For our analysis, the daily returns were collected over a period from October 11, 2022, to January 19, 2023, covering 50 days before and after the release. Stocks with prolonged missing values were removed, resulting in a dataset with $d = 82$ dimensions, and group sizes of $m = 34$ and $n = 32$ for the two comparison groups.

For the proposed MATES method, we use the same configuration as in the simulation studies to construct the test statistic.
We compare our method with all the comparison methods in the simulation. 
The $p$-values of these tests (except for aMMD) are presented in \Cref{tab:SP100}, with those smaller than 0.05 highlighted in bold. 
For this significance level, the aMMD test uses a threshold of $0.017$ to adjust for multiple comparisons, and with the calculated minimum $p$-value of $0.049 > 0.017$, it fails to reject the null hypothesis. 
From the result, only the proposed method, BD, and GPK, can detect the pattern difference before and after the ChatGPT release at a significance level of $0.05$. However, the proposed method is the only one that rejects the null hypothesis at a significance level of $0.01$. This indicates that the proposed method may be more powerful than the other methods in this comparison.

\begin{table}[!htpb]
\caption{The $p$-values of the tests for the S\&P100 stock return.}
\label{tab:SP100}
\begin{center}
\resizebox{\textwidth}{!}{%
\begin{tabular}{cccccccccccc}
\toprule
 MATES& CM& GET& BD& GED& RF& MT& GPK& RISE& MMD& xMMD& mMMD      \\
\midrule
 $<$\textbf{0.001} & 0.786 & 0.071 & \textbf{0.025} & 0.148 & 0.208 & 0.065 & \textbf{0.039} & 0.08 & 0.213 & 0.304 & 0.637\\ 
\bottomrule   
\end{tabular}
}
\end{center}
\end{table}

\begin{figure}
    \centering
    \includegraphics[width=0.8\linewidth]{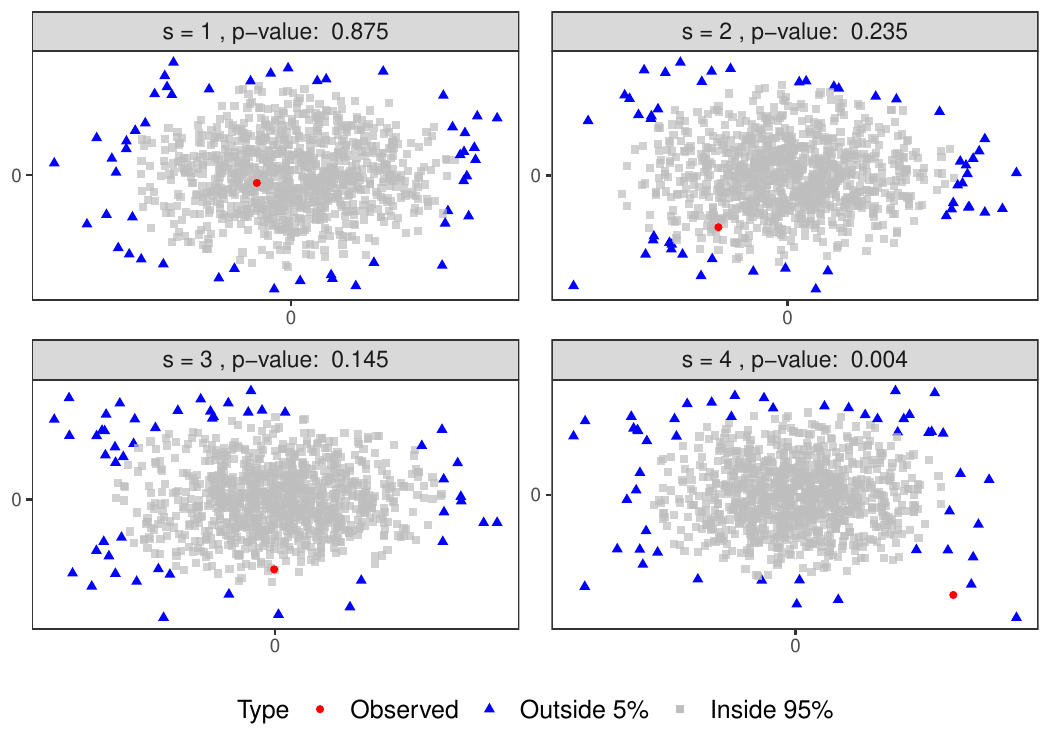}
    \caption{Scatterplots of observed and permuted $(U_w^{(s)}, U_{\text{diff}}^{(s)})$ with $1000$ permutations for the S\&P100 data.}
    \label{fig:SP100}
\end{figure}

To further explore the underlying patterns and confirm our conclusion, we make the scatter plot of observed statistics $(U_{w}^{(s)}, U_{\text{diff}}^{(s)})$ for $s = 1,2,3,4$ in \Cref{fig:SP100}. The red dots correspond to the observed test statistic, while the gray and blue dots represent $1000$ permuted samples. The $p$-values associated with each view are obtained in the following way. First, the single-view test statistic based on the $s$th moment can be represented by $T'_s = \bu_s \trans \bSigma_{\bu}^{-1} \bu_s$, where $\bu_s = (U_x^{(s)} - \mu_x^{(s)}, U_y^{(s)} - \mu_y^{(s)})\trans$ for $s=1,2,3,4$ and $\bSigma_{\bu} = \Cov(\bu_s)$. By \Cref{thm:limitDist}, each $T'_s$ converges in distribution to $\chi^2_2$ under the permutation null, so the $p$-values are obtained from the tail probability of a $\chi^2_2$ distribution.

From the figure we can see that when considering only lower moments, i.e. the first three views, the observed data point is indistinguishable from the permutation samples. When considering the difference in the fourth moment (via the fourth view), the $p$-value becomes $0.004$ and we reject the null at the $0.05$ significant level. This indicates that the distribution differences are inherited in the higher moments for the S\&P 100 data. In fact, MATES provides a $p$-value of less than $0.001$, indicating even stronger evidence against the null hypothesis when aggregating available information from all four moments. 

\section{Discussion}\label{sec:discussion}
We introduce a novel graph-based test statistic that aggregates information from multiple views, with each view summarizing a distinct characteristic of the two-sample data. The general framework can accommodate a variety of similarity measures, similarity graphs, and edge weights, in the construction of test. It enables the combination of popular two-sample test methods such as graph-, kernel-, distance-, and rank-based methods that use similarity graphs, kernel values, pairwise distance, and ranks, respectively, to define individual views, which are subsequently combined to produce a more powerful test statistic. 

This approach addresses the limitations of existing methods in detecting higher-order distributional differences, and offers robust and stable performance across various dimensions and distributional scenarios. The theoretical analysis establishes a distribution-free limiting distribution under the null hypothesis, thereby eliminating the computational burden associated with resampling methods like permutation or bootstrap.

There are several directions that can extend the utility and applicability of the proposed methodology. In our numerical studies and real data applications, we focus on the first to fourth moments of the data to construct multiple views. While it may seem advantageous to include as many moments as possible to enhance the power of the test, doing so can increase the computational complexity without substantial gains. A direction for future work is the development of data-adaptive procedures to select a parsimonious set of views that can be equally powerful. Additionally, extending the framework to consider a continuum of views may be of more interest for some settings where continuous shifts in distributions are more informative.

The concept of view aggregation underlying MATES can be generalized to other statistical tasks as well. For example, a natural extension of this framework is change-point detection, where the goal is to identify shifts in distribution over time. By aggregating information from multiple views across temporal windows, the framework could detect subtle changes in dynamic systems. Similarly, integrating the view aggregation concept into classification or clustering could leverage distributional differences to enhance model performance.

\bibliographystyle{apalike}
\spacingset{1.45}

\end{document}